# First Demonstration of 25λ × 10 Gb/s C+L Band Classical / DV-QKD Co-Existence Over Single Bidirectional Fiber Link

Florian Honz, *Student Member, IEEE*, Florian Prawits, Obada Alia, Hesham Sakr, Thomas Bradley, *Member, IEEE*, Cong Zhang, Radan Slavík, *Senior Member, IEEE*, Francesco Poletti, *Member, IEEE*, George Kanellos, Reza Nejabati, *Senior Member, IEEE*, Philip Walther, Dimitra Simeonidou, *Fellow, IEEE*, Hannes Hübel and Bernhard Schrenk, *Member, IEEE*

*Abstract*—As quantum key distribution has reached the maturity level for practical deployment, questions about the co-integration with existing classical communication systems are of utmost importance. To this end we demonstrate how the co-propagation of classical and quantum signals can benefit from the development of novel hollow-core fibers. We demonstrate a secure key rate of 330 bit/s for a quantum channel at 1538 nm in the presence of 25 × 10 Gb/s classical channels, transmitted at an aggregated launch power of 12 dBm, spanning over the C+L-band in the same hollow-core fiber link. Furthermore, we show the co-integration of the classical key-distillation channel onto this fiber link, turning it into a bidirectional fiber link and thereby mitigating the need for multiple fibers. We believe this to be an important step towards the deployment and integration of hollow-core fibers together with DV-QKD for the inherently secure telecom network of the future.

## I. INTRODUCTION

QUANTUM computing is posing a permanent threat to the security of our currently employed communication protocols. Luckily it can be countered by employing information-theoretic secure quantum key distribution (QKD) protocols to secure the transmitted and stored data. In the recent years commercial QKD solutions have been blooming, advancing QKD from lab-scale experiments to field-installed products which mainly focus on securing point-to-point links. However, the practical network integration of QKD together with classical channels, featuring a ~90 dB higher per-channel launch power, remains a challenge. Therefore, for most QKD demonstrations the necessity of a dedicated dark fiber link for the QKD channel arose, in order to separate the QKD and the classical channels spatially. In fiber-scarce environments as well as under stringent operational expenditures such a dark fiber, dedicated to the QKD channel, is not a viable option. In view of this, recent works [1-35] (Fig. 1) have started to investigate the influence of the co-propagation of the classical channels over the same fiber link as the QKD signal as well as the robustness of the QKD link under these conditions. This led to the identification of in-band crosstalk noise due to Raman scattering as the primary constraint, which's influence is of such high significance due to the high power difference between the classical and quantum power levels [36]. These conditions would therefore favor the deployment of continuous-variable (CV) QKD (● in Fig. 1) due to its noise filtering capabilities thanks to the employed coherent reception methodology [23-35]. However, CV-QKD is limited to short reach applications and the generation of secure keys cannot be guaranteed when high optical budgets are to be overcome. An alternative approach to CV-QKD would be to dedicate the remote O-band (□ in Fig. 1) [12-22] to loss-tolerant discrete-variable (DV) QKD [8], enhancing the robustness of the system to the noise induced by the classical channels. However, the higher fiber loss in the O-band is limiting the reach of such a hybrid

This work received funding from the EU Horizon-2020 program (grant 820474) and the Austrian Research Promotion Agency FFG (grant 881112).

F. Honz, F. Prawits, H. Hübel and B. Schrenk are with the AIT Austrian Institute of Technology, Center for Digital Safety & Security, Giefinggasse 4, 1210 Vienna, Austria (e-mail: florian.honz@ait.ac.at).

O. Alia, G. Kanellos, R. Nejabati and D. Simeonidou are with the High Performance Network Group, School of Computer Science, Electrical and Electronic Engineering and Engineering Maths (SCEEM), University of Bristol, BS8 1TH Bristol, U.K. (e-mail: obada.alia@bristol.ac.uk).

H. Sakr, T. Bradley, C. Zhang, R. Slavik and F. Poletti are with the Optoelectronics Research Centre, University of Southampton, Southampton, SO17 1BJ, U.K. (e-mail: frap@orc.soton.ac.uk).

P. Walther is with the University of Vienna, Faculty of Physics, Vienna Center for Quantum Science and Technology (VCQ), Boltzmanngasse 5, 1090 Vienna, Austria (e-mail: philip.walther@univie.ac.at).

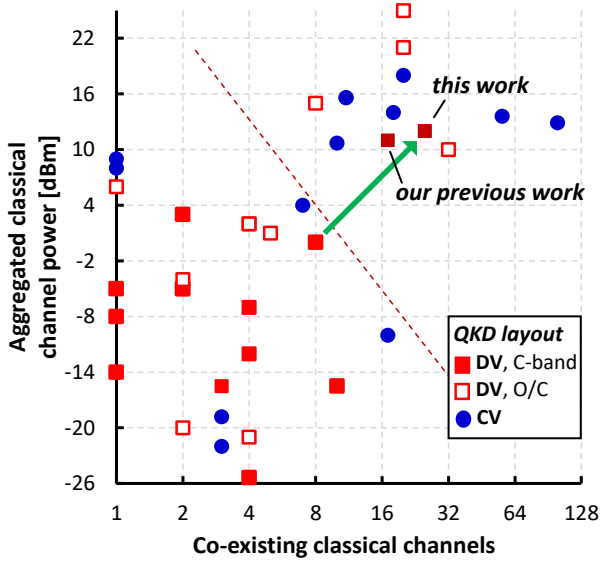

Fig. 1. QKD co-existence demonstrations [1-35, 38].

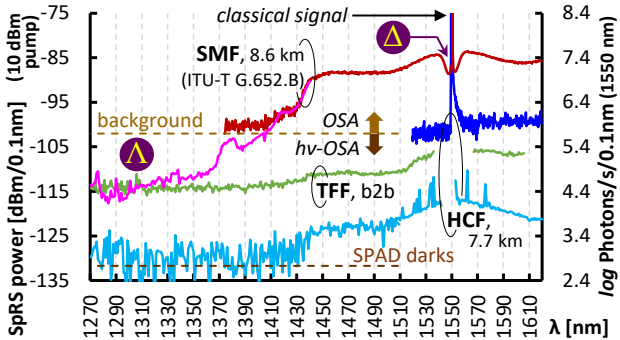

Fig. 2. Spontaneous Raman scattering spectra of SMF, HCF and TFF. In the SMF, the DV-QKD channel needs to be allocated in the O-band ($\Lambda$) or in the notch of the Raman spectrum ($\Delta$) in order to ensure minimal degradation due to Raman noise.

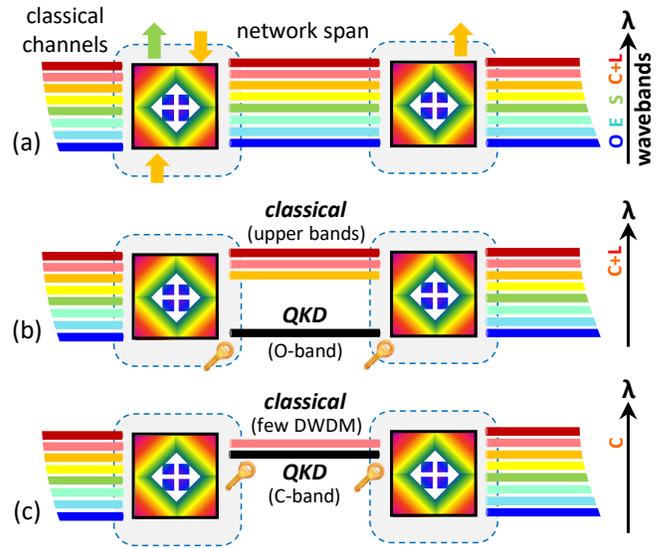

Fig. 3. QKD integration challenge for metro-core links: (a) classical ultra-wideband system utilizing the full fiber spectrum, (b) reduction of the exploitable wavebands due to the use of two remote wavebands for the QKD and classical signals and (c) restriction to a few DWDM channels when using the same waveband for the QKD and the classical signals.

O/C-band QKD/classical system. On top of this, it is hardly compatible with the spectral expansion envisioned for ultra-wideband transmission schemes [37].

In this work, we expand our previous experimental investigation of the C-band classical / 1538 nm DV-QKD co-existence [38], employing the same metro-scale hollow-core fiber (HCF) link featuring reduced Raman scattering. We are not only expanding the range of the classical channels from the C-band to the L-band, but we also show, for the first time to our best knowledge, the simultaneous co-propagation of the classical key-distillation channels on the same fiber. We accomplish secure-key generation at 330 bit/s while loading the link with 25 data channels which aggregate at a power of 12 dBm. This corresponds to an improvement of 9 dB in classical co-existence power while at the same time the number of co-propagating channels is increased by a factor of 2.5 compared to previous experiments [11]. Furthermore, we demonstrate long-term stability of our bidirectional link, using only the HCF for simultaneous transmission of the classical load channels, the quantum channel, the frame synchronization channel and the bidirectional key-distillation channels.

The paper is organized as follows: Section II covers the impairments of the co-propagating classical channels and the influence of their Raman noise on the allocation and performance of the quantum channel. The experimental setup is discussed in Section III. Section IV then focuses on the performance of the system regarding the achievable secure key rates (SKR) as well as the performance of the classical channels. Finally, Section V summarizes our work and concludes with an outlook on possible further improvements and applications.

## II. Raman Noise and its Influence on the Spectral Allocation of the Quantum Channel

Strong classical channels propagating in a medium cause Raman scattering due to the interaction of the electromagnetic wave with the molecules of the medium. In a standard single-mode fiber (SMF) compatible with ITU-T G.652.B, this results in inevitable in-band Raman noise contaminating the spectrum when classical and quantum channels with huge power differences of typically nine orders of magnitude co-propagate on the same fiber. As can be seen from the measurements with a classical optical spectrum analyzer (OSA) and a single-photon ($h\nu$) resolving OSA, the spontaneous Raman scattering (SpRS) for a 1550-nm pump beam in a SMF is spectrally broad (Fig. 2). However, exploiting the O-band, where the noise contribution of the SpRS tails finally falls below $10^5$ photons/s/100pm ($\Lambda$ in Fig. 2), appears to be feasible for QKD integration, since this value is compatible with DV-QKD systems [12-22]. Considering C-band QKD integration, the magnitude of Raman noise in the C-band, as well as its adjacent wavebands, does not allow for a wideband layout of the classical channels. This is evidenced by previous works, such as [8], where only a very limited number of classical channels co-exist with DV-QKD in the C-band (■ in Fig. 1). These classical channels are spectrally allocated very closely to the DV quantum channel which is necessary to ensure a weak Raman noise contribution (Δ in Fig. 2). Additionally, the attenuation of the classical channels becomes a necessity to use the same waveband for classical and quantum signals.

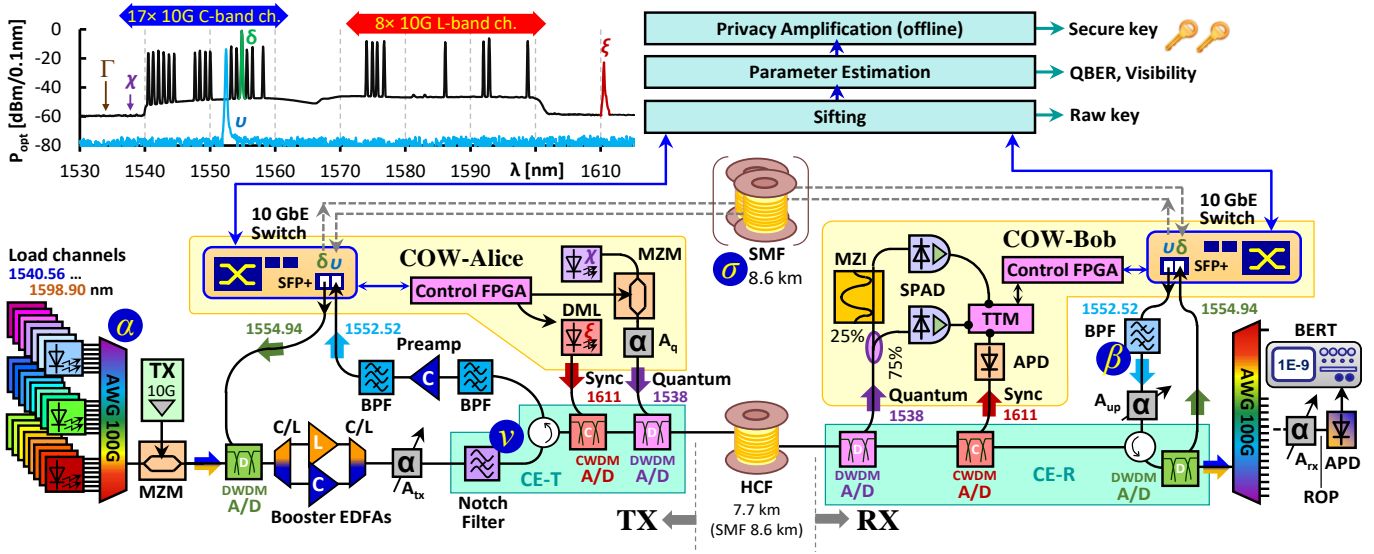

Fig. 4. Experimental Setup for the joint, bidirectional classical/quantum C+L band transmission. The spectral inset presents the transmitted spectrum, including the down- and uplink ($\delta$, $\nu$) as well as the synchronization channel ($\xi$). The quantum channel ($\chi$) is not visible due to the noise background originating from the OSA grating ($\Gamma$). $\alpha$ denotes the AWG combining the classical channels, $\beta$ the filter for the uplink channel, $\nu$ the uplink channel and $\sigma$ the fiber pair that was initially used for the classical key-distillation channels, before these channels had been transmitted over the HCF.

Both, the restrictions in terms of classical spectral bandwidth and in classical signal launch power impose a roadblock for the seamless introduction of QKD in metro-core networks. This practical QKD integration challenge is discussed in Fig. 3. Ultra-wideband systems aim to expand the utilized fiber spectrum for classical data transmission from the O- to the L-band, as highlighted in Fig. 3a. The practical integration of DV-QKD has been targeted following two options, each of them resulting in an unfavorable trade-off with co-existing classical traffic. One approach is the use of two remote wavebands, such as shown for O-band QKD and C-band classical channels in Fig. 3b. With this, however, the vast fiber spectrum cannot be exploited anymore. Alternatively, the classical channels can be spectrally allocated very closely to the QKD channel, within the same waveband (Fig. 3c). In this case, however, the use of wavelength division multiplexing (WDM) is restricted to very few dense WDM (DWDM) channels. This, however, then stands against the main aim of metro and core networks to bridge their network nodes with the highest possible capacity.

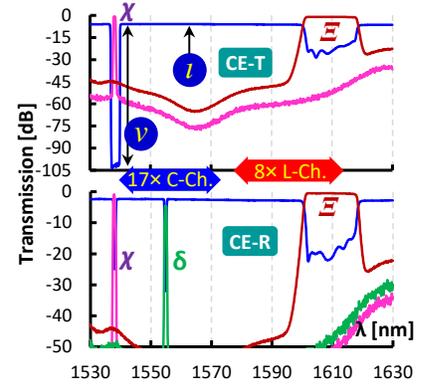

Fig. 5. Transmission spectra of the co-existence elements at the transmitter (top) and the receiver (bottom), with the express feed-through path ($\iota$) for the classical channels, the notch ($\nu$) for the quantum channel ($\chi$) and the add/drop for the downlink ($\delta$) and QKD synchronization ($\Xi$) channels.

To enable wideband use of the C+L bands for classical communication while also using a C-band DV-QKD channel, we build on the initial findings from [11], where a HCF with an air-filled core is used. The advantage of the HCF resides in the elimination of Raman scattering and four-wave mixing due to the air-filled core. The HCF sample which we used for our experiment had an overall length of 7.7 km, an attenuation of 9.1 dB and was spliced to SMF-28 fiber pigtails using mode field adapters. It is comprised of two spliced sections of hollow-core Nested Antiresonant Nodeless Fibers (NANFs). The sections are from different draws and feature slightly different core diameters (35.6 μm and 35.9 μm), membrane thicknesses and losses, as described in detail in [39]. Due to their design, NANFs also feature a reduced inter-modal interference and several 100 nm broad transmission windows. As can be seen in Fig. 2, the SpRS of this 7.7 km HCF sample is reduced by ~35 dB compared to the SMF, allowing us to use the whole C+L-bands for classical data transmission.

We would like to stress that the newest HCF fibers now reach losses as low as 0.174 dB/km [40]. This corresponds to a total loss of 1.3 dB for a link length of 7.7 km and is already lower than the loss of an ITU-T G.652.B compliant SMF at 1550 nm.

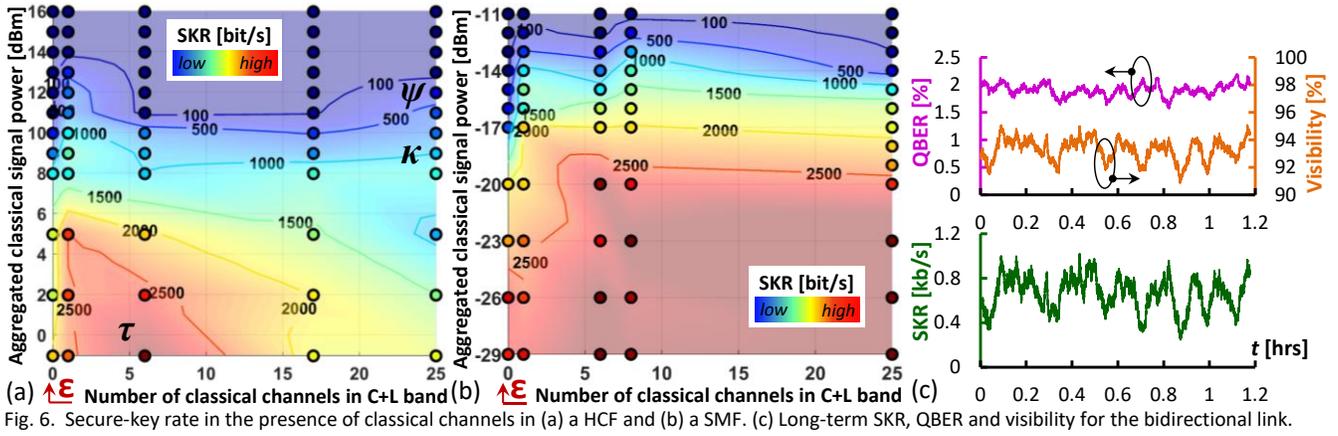

Fig. 6. Secure-key rate in the presence of classical channels in (a) a HCF and (b) a SMF. (c) Long-term SKR, QBER and visibility for the bidirectional link.

### III. EXPERIMENTAL SETUP AND SPECTRAL LAYOUT

The experimental setup is presented in Fig. 4. The QKD channel was implemented based on the coherent one-way (COW) protocol described in detail in [41]. At the transmitter side we used a laser at 1538 nm ($\chi$) as our light source for the quantum channel. The output of this laser was then pulsed at a rate of 1 GHz using a LiNbO$_3$ Mach-Zehnder-Modulator (MZM), this way forming a coherent pulse train, and afterwards attenuated ($A_q$) to an average photon number of $\mu$ = 0.1 photons/pulse. The transmitted coherent pulse train consisted of individual frames with a length of 64 time slots, each of them with a length of one nanosecond, and the overall frame period was 1024 ns. Additionally at the beginning of each frame a synchronization pulse at 1611 nm ($\xi$), generated by a directly modulated laser, was transmitted to act as the local time reference for the QKD receiver. At the receiver side we used two free-running InGaAs single-photon avalanche photodiodes (SPAD): One to directly detect the incoming photons and their time of arrival (75% branch) and one to detect the phase between two consecutive pulses (25% branch) using a Mach-Zehnder interferometer (MZI), which helps to identify decoy states. The detectors had an efficiency of 10% and a dark count rate of 620 counts/s. The frame synchronization pulses were detected using a classical APD and recorded, together with the SPAD signals, with a time-tagging module (TTM). The whole QKD system was operated via FPGA-based controllers. Those controllers performed pattern generation, synchronization and registration of the detection events and fed this information to the key-distillation stack, consisting of (*i*) real-time raw key sifting, (*ii*) a real-time parameter estimation to calculate the quantum bit error ratio (QBER) and the visibility, and (*iii*) offline privacy amplification, after which we were eventually able to generate a secure key.

The classical communication channel over which the key distillation was implemented is spanned between two Ethernet switches equipped with SFP+ optics capable of operating at 10 Gb/s. The respective 10GbE downlink channel from Alice to Bob was placed at 1554.94 nm ($\delta$), while the uplink channel was located at 1552.52 nm ($\nu$). For the experiment employing 17 C-band channels, described in our previous paper [38], we had placed this bidirectional classical channel on a separate fiber pair ($\sigma$). In the present experiment we retained this configuration at first and expanded the classical optical spectrum to 25 C+L band channels. In a second step, we also integrated the bidirectional key-distillation channel on the HCF link, using circulators to implement the required directional split. To reduce any backscattering from the 10GbE uplink channel into the quantum channel at the co-existence elements of the receiving end (CE-R), which would cause a degradation for the quantum signal, we filtered the 10GbE uplink channel ($\beta$) and then attenuated it ($A_{up}$) to optimize the SKR. Due to the unavailability of sensitive SFP+ modules with APD-based detectors, the use of a filtered optical preamplifier at the uplink receiver was necessary in order to enhance the level and the signal-to-noise ratio of the received 10GbE uplink signal.

To study the C+L band co-propagation with the DV-QKD signal, we loaded the HCF link with 25 classical channels along the ITU-T grid, ranging from 1540.56 nm to 1598.90 nm. These channels were modulated at 10 Gb/s with pseudo-random data (PRBS) and multiplexed using an arrayed waveguide grating (AWG, $\alpha$). Due to the co-existence of the classical and the QKD channels in the HCF, the necessity for spectral conditioning of the signal spectrum arises. To this regard the AWG serves as a filter for the spontaneous emission tails of the laser emission, whereas the SFP+ 10GbE downlink module is filtered by the corresponding DWDM add/drop element. After this initial filtering, the downstream signals are boosted using a tandem of C- and L-band EDFAs. After levelling up the signal power, an attenuator ($A_{tx}$) is used to control the co-existence power of the classical channels. This allows us to investigate the co-existence limits for secure-key generation. A notch filter with a rejection of >95 dB (Fig. 5, $\nu$), comprised of a cascade of thin-film filters, then suppresses the amplified spontaneous emission (ASE) noise of the EDFAs and cleans out the quantum channel at 1538 nm. For the bidirectional setup used for the long-term measurements, we exchanged the classical 10 Gb/s modulated channels at 1554.94 nm and 1552.52 nm with our SFP+ sourced

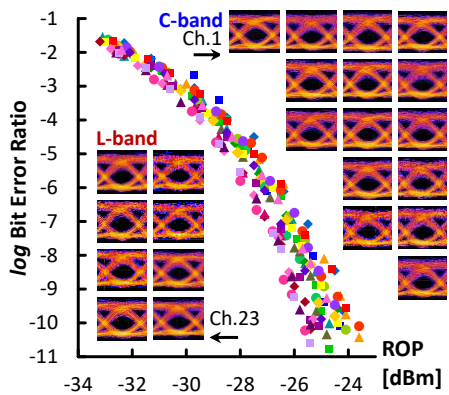

Fig. 7. BER and eye diagrams for all PRBS channels.

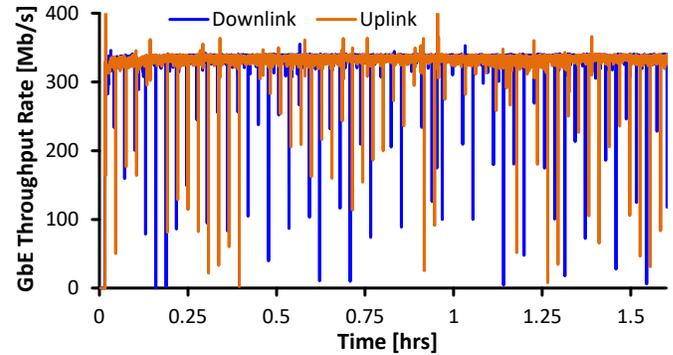

Fig. 8. Long-time GbE throughput rate for 1 Gb/s.

10GbE channels, so that the HCF link was once again loaded with a total of 25 classical channels.

To further reduce any impairments of the classical channels on the quantum signal, the latter was multiplexed lastly to the HCF. This also ensures that the multiplexing loss for the quantum signal remains minimal with 0.76 dB (Fig. 5, $\chi$), while the loss of the classical channels due to notching and express feed-through paths amounts to a total of 6.3 dB at the transmitter side (CE-T, $\iota$). The complete spectrum of the optical signals which are being transmitted through the HCF is reported as an inset in Fig. 4. The classical signals induce an additional scattering at the OSA grating, which provides an artificial noise background ($\Gamma$). Therefore, the quantum channel as well as the notched transmission window ($\chi$) cannot be resolved in detail due to the limited dynamic range for the spectral acquisition.

After the transmission through the HCF, the receiver-side co-existence element (CE-R) firstly drops the quantum signal, ensuring a minimal demultiplexing loss of 0.74 dB. The demultiplexing procedure is continued by the synchronization channel ($\Xi$), the downlink SFP+ channel ($\delta$) and the classical channels, which experience a total loss of 2.5 dB. The classical channels are further demultiplexed by an AWG and detected by an APD receiver. The quality of the signals is then evaluated via bit error ratio (BER) measurements as a function of the received optical power (ROP), for which a variable optical attenuator ($A_{rx}$) was employed.

In order to be able to directly compare the system performance to an SMF-based transmission scenario, we finally exchanged the HCF by a SMF with a length of 8.6 km.

IV. SYSTEM PERFORMANCE UNDER CO-EXISTENCE

*A. QKD Performance – Secure-key rate under link load*

As already mentioned, we firstly placed the 10GbE key-distillation channel on a separate fiber pair to evaluate the SKR as a function of the number of classical channels and their aggregated classical power. This was done in order to reduce long set-up times when continuously altering the loss conditions for the classical channels, which would require a re-synchronization of the involved 10GbE switches.

Figure 6a reports the accomplished SKRs as function of the aggregated classical launch power and the number of classical channels. Since the spectral Raman scattering profile of the HCF is flat and low (Fig. 2), the achievable SKR is mainly determined by the aggregated power of the classical channels. As can be seen in Fig. 6a and expected from the flat Raman noise spectrum, the SKR is widely independent of the spectral layout of the classical channels. We achieved a SKR of 330 bit/s ($\psi$) for 25 co-propagating classical channels with an aggregated power of 12 dBm. This clearly shows the robustness of DV-QKD to a DWDM feed of classical data channels spanning over 58.3 nm in the C+L bands in case of transmission over a HCF. Furthermore, the classical channels are not restricted to a narrow (few nm) spectral slice close to the quantum channel anymore ($\Delta$ in Fig. 2), like it would be required in case of a SMF. This means that DV-QKD transmission becomes compatible with WDM transmission – a prerequisite for the practical introduction of QKD in metro and core networks for which the classical channel layout cannot be altered.

For a weaker launch power of 9 dBm, we achieve a SKR of 1 kb/s at a QBER of 1.31 % ($\kappa$). This is sufficiently high to secure the whole link capacity of 25 × 10 Gb/s according to the NIST recommendation for AES key renewal, which requires one 256-bit long AES key for every data chunk of 64 Gbyte. If we only transmit the six C-band channels closest to the quantum channel, ranging from 1540.56 nm to 1544.53 nm, we can reach a SKR of up to 3.1 kb/s ($\tau$) at an aggregated classical launch power of -1 dBm due to the low intrinsic QBER of 0.67 % of the COW-QKD system. This demarks the highest key-rate we could achieve for any number of classical channels combined with the quantum channel.

Figure 6b shows the SKR performance with a SMF instead of the HCF for a direct comparison. We were not able to generate a SKR for a co-existence power larger than -13 dBm, where we eventually reach a SKR of 107 b/s in presence of all classical channels. To obtain a SKR of 1 kb/s, we were required to attenuate the classical channels to an aggregated launch power of -15 dBm. Translating the higher permissible signal launch power for a HCF to an unallocated optical budget, we see that the link budget can be extended by an enormous amount of 24 dB with respect to SMF transmission.

For both fiber types a drop in the SKR to 1.5 ... 2 kb/s and 2 ... 2.5 kb/s can be noticed when no classical channels are active ($\varepsilon$ in Figs. 6a, b). We attribute this artifact to the considerably elevated ASE background of the booster EDFAs under no-load conditions, which enhances the contamination of the quantum channel due to the limited extinction of the employed notch filter.

In order to elaborate on the limits of co-existence, we investigated a back-to-back configuration without fiber span. Here we could still measure a low background noise, whose origin we attribute to the SMF-pigtailed thin-film WDM filters at the transmitter-side co-existence element CE-T. Figure 2 includes the SpRs for one of the employed thin-film filter (TFF) elements. It shows a clearly stronger SpRS response than the HCF, even after correction of the HCF transmission loss. This confirms the existence of an upper bound for the permissible launch power of the classical channels due to filter elements with SMF pigtails and proves that the link-level Raman scattering cannot be fully mitigated but is rather brought to a level of fiber-optic pigtails. Further studies on purely HCF-based filter elements would be required to investigate higher classical power levels in co-existence with QKD multiplexing in the same waveband.

*B. QKD Performance – Long-term stability*

We then switched to the bidirectional setup, integrating the 10GbE key-distillation channels ($\delta$, $v$) bidirectionally with the HCF link and adjusting the launch power for the 24 classical downlink channels to 9 dBm. The launch power of the 10GbE uplink channel was set to -12.5 dBm. We performed a long-term measurement over 70 minutes (Fig. 6c) and the resulting average secure-key generation ratio, which is tightly linked to the visibility, was $2.1\times10^{-6}$, corresponding to an average SKR of 660 bit/s. Furthermore, the QKD performance remained stable for the entire duration of the measurement, yielding a QBER of 1.91% (3$\sigma$ = 0.313%) and a visibility of 93.3% (3$\sigma$ = 2.33%).

*C. Classical Data Transmission Performance*

To conclude our investigations, we finally evaluated the performance of our classical channels. We were able to obtain a sensitivity better than -23.6 dBm at a BER of $10^{-10}$. The spread among all channels between the best and the worst ROP at this BER threshold was 2.5 dB. Furthermore, all eye diagrams of the load channels were clearly open, as is evidenced in Fig. 7.

Additionally, we evaluated the real-time GbE throughput rate of our bidirectional key-distillation channel over the 10GbE connection. It yields an average data rate of ~330 Mb/s for the down- and uplink directions (Fig. 8). Although there are periodic drops in the data rate, which we attribute to the chunk-wise data emulation and buffering, there was no temporary link interruption observed due to the low upstream launch power, rendering the chosen set-point stable.

V. CONCLUSION

We successfully demonstrated secure-key generation on a 1538-nm quantum channel integrated with a bidirectional key-distillation channel, a L-band synchronization channel and 23 co-transmitted classical DWDM load channels in the C+L-bands from 1540.56 nm to 1598.89 nm on the same hollow-core fiber link. We were able to accomplish a SKR of 330 bit/s for 25 classical channels at an aggregated classical signal power of 12 dBm for a unidirectional fiber link, while further obtaining an average key-rate of 660 bit/s for a classical launch power of 9 dBm for a bidirectional link. By expanding the optical spectrum dedicated to the classical channels to the C+L-bands, we achieve a 25-dB improvement in the product of classical power × optical bandwidth compared to earlier DV-QKD works investigating the co-existence of QKD and classical channels in the C-band. This now enables the integration of DV-QKD in high-capacity, long-range metro-core links without restricting the use of WDM for classical channel transmission any longer.

Furthermore, the need for upgrading fiber-optic components to unleash even higher classical power levels of 20 dBm and beyond arises: We found thin-film add/drop multiplexers, which combine the quantum channel and the 90-dB stronger classical signals, to be the performance limiting factor due to their SMF pigtails. Towards this direction, we expect the development of hollow-core WDM components together with sub-DWDM filtering at the quantum channel as the next steps to allow further scaling of QKD integration in optical telecommunication networks.

> REPLACE THIS LINE WITH YOUR PAPER IDENTIFICATION NUMBER (DOUBLE-CLICK HERE TO EDIT) < 8[28] T. A. Eriksson, T. Hirano, M. Ono, M. Fujiwara, R. Namiki, K. Yoshino, A. Tajima, M. Takeoka, and M. Sasaki, "Coexistence of continuous variable quantum key distribution and 7×12.5 Gbit/s classical channels," in *Proc. IEEE Photon. Soc. Summer Top. Meeting Ser.*, Waikoloa Village, HI, USA, July 2018, pp. 71–72, DOI: 10.1109/PHOSST.2018.8456709.

[29] T. A. Eriksson, T. Hirano, G. Rademacher, B. J. Puttnam, R. S. Luís, M. Fujiwara, R. Namiki, Y. Awaji, M. Takeoka, N. Wada, and M. Sasaki, "Joint propagation of continuous variable quantum key distribution and 18 × 24.5 Gbaud PM-16QAM channels," in *Proc. Eur. Conf. Opt. Commun.*, Rome, Italy, Sept. 2018, pp. 1-3, DOI: 10.1109/ECOC.2018.8535421.

[30] T. A. Eriksson, T. Hirano, B. J. Puttnam, G. Rademacher, R. S. Luís, M. Fujiwara, R. Namiki, Y. Awaji, M. Takeoka, N. Wada, and M. Sasaki, "Wavelength division multiplexing of continuous variable quantum key distribution and 18.3 Tbit/s data channels," *Nat. Commun.*, vol. 2, no. 9, Jan. 2019, DOI: 10.1038/s42005-018-0105-5.

[31] S. Kleis, J. Steinmayer, R. H. Derksen, and C. G. Schaeffer, "Experimental investigation of heterodyne quantum key distribution in the S-band embedded in a commercial DWDM system," in *Proc. Opt. Fiber Commun. Conf.*, San Diego, CA, USA, Mar. 2019, pp. 1-3, DOI: 10.1364/OFC.2019.Th1J.3.

[32] A. Aguado, V. López, D. López, M. Peev, A. Poppe, A. Pastor, J. Folgueira, and V. Martín, "The engineering of software-defined quantum key distribution networks," in *IEEE Commun. Mag.*, vol. 57, no. 7, pp. 20–26, July 2019, DOI: 10.1109/MCOM.2019.1800763.

[33] D. Milovančev, N. Vokić, F. Laudenbach, C. Pacher, H. Hübel, and B. Schrenk, "Spectrally-shaped continuous-variable QKD operating at 500 MHz over an optical pipe lit by 11 DWDM channels," in *Proc. Opt. Fiber Commun. Conf.*, San Diego, CA, USA, Mar. 2020, pp. 1-3, DOI: 10.1364/OFC.2020.T3D.4.

[34] R. Valivarthi, S. Etcheverry, J. Aldama, F. Zwiehoff, and V. Pruneri, "Plug-and-play continuous-variable quantum key distribution for metropolitan networks," *Opt. Express*, vol. 28, no. 10, pp. 14547–14559, May 2020, DOI: 10.1364/OE.391491.

[35] D. Milovančev, N. Vokić, F. Laudenbach, C. Pacher, H. Hübel, and B. Schrenk, "High Rate CV-QKD Secured Mobile WDM Fronthaul for Dense 5G Radio Networks," *J. Light. Technol.*, vol. 39, no. 11, pp. 3445-3457, June 2021, DOI: 10.1109/JLT.2021.3068963.

[36] S. Aleksic, F. Hipp, D. Winkler, A. Poppe, B. Schrenk, and G. Franzl, "Perspectives and limitations of QKD integration in metropolitan area networks," *Opt. Express*, vol. 23, no. 8, pp. 10359-10373, Apr. 2015, DOI: 10.1364/OE.23.010359.

[37] A. Ferrari, A. Napoli, J. K. Fischer, N. Costa, A. D'Amico, J. Pedro, W. Forysiak, E. Pincemin, A. Lord, A. Stavdas, J. P. F.-P. Gimenez, G. Roelkens, N. Calabretta, S. Abrate, B. Sommerkorn-Krombholz, and V. Curri, "Assessment on the Achievable Throughput of Multi-Band ITU-T G.652.D Fiber Transmission Systems," *J. Light. Technol.*, vol. 38, no. 16, pp. 4279-4291, Aug. 2020, DOI: 10.1109/JLT.2020.2989620.

[38] F. Honz, F. Prawits, O. Alia, H. Sakr, T. D. Bradley, C. Zhang, R. Slavík, F. Poletti, G. T. Kanellos, R. Nejabati, P. Walther, D. Simeonidou, H. Hübel, and B. Schrenk, "Demonstration of 17λ x 10 Gb/s C-Band Classical / DV-QKD Co-Existence Over Hollow-Core Fiber Link," in *Proc. Eur. Conf. Opt. Commun.*, Basel, Switzerland, Sept. 2022, pp. 1-4, paper Th1G.3.

[39] A. Nespola, S. Straullu, T. D. Bradley, K. Harrington, H. Sakr, G. T. Jasion, E. N. Fokoua, Y. Jung, Y. Chen, J. R. Hayes, F. Forghieri, D. J. Richardson, F. Poletti, G. Bosco, and P. Poggiolini, "Transmission of 61 C-Band Channels Over Record Distance of Hollow-Core-Fiber With L-Band Interferers," *J. Light. Technol.*, vol. 39, no. 3, pp. 813-820, Feb. 2021, DOI: 10.1109/JLT.2020.3047670.

[40] G. T. Jasion, H. Sakr, J. R. Hayes, S. R. Sandoghchi, L. Hooper, E. N. Fokoua, A. Saljoghei, H. C. Mulvad, M. Alonso, A. Taranta, T. D. Bradley, I. A. Davidson, Y. Chen, D. J. Richardson and F. Poletti, "0.174 dB/km Hollow Core Double Nested Antiresonant Nodeless Fiber (DNANF)," in *Proc. Opt. Fiber Commun. Conf.*, San Diego, CA, USA, Mar. 2022, pp. 1-3, DOI: 10.1364/OFC.2022.Th4C.7.

[41] D. Stucki, N. Brunner, N. Gisin, V. Scarani, and H. Zbinden, "Fast and simple one-way quantum key distribution," *Appl. Phys. Lett.*, vol. 87, no. 19, pp. 194108, Nov. 2005, DOI: 10.1063/1.2126792.